\documentclass[10pt,twocolumn,letterpaper]{article}

\usepackage{cvpr}
\usepackage{times}
\usepackage{epsfig}
\usepackage[numbers]{natbib}
\usepackage{caption}
\usepackage{graphicx}
\usepackage{amsmath}
\usepackage{amssymb}
\usepackage{multirow}
\usepackage[breaklinks=true,bookmarks=false]{hyperref}

\cvprfinalcopy
\begin{document}

\title{Free-text Keystroke Authentication using Transformers: A Comparative Study of Architectures and Loss Functions}

\author{Saleh Momeni\\
\small School of Mathematics, Statistics, and Computer Science\\
\small University of Tehran, Tehran, Iran\\
{\tt\small saleh.momeni@ut.ac.ir}
\and
Bagher BabaAli*\\
\small School of Mathematics, Statistics, and Computer Science\\
\small University of Tehran, Tehran, Iran\\
{\tt\small babaali@ut.ac.ir}
}

\maketitle

\begin{abstract}
Keystroke biometrics is a promising approach for user identification and verification, leveraging the unique patterns in individuals' typing behavior. In this paper, we propose a Transformer-based network that employs self-attention to extract informative features from keystroke sequences, surpassing the performance of traditional Recurrent Neural Networks. We explore two distinct architectures, namely bi-encoder and cross-encoder, and compare their effectiveness in keystroke authentication. Furthermore, we investigate different loss functions, including triplet, batch-all triplet, and WDCL loss, along with various distance metrics such as Euclidean, Manhattan, and cosine distances. These experiments allow us to optimize the training process and enhance the performance of our model. To evaluate our proposed model, we employ the Aalto desktop keystroke dataset. The results demonstrate that the bi-encoder architecture with batch-all triplet loss and cosine distance achieves the best performance, yielding an exceptional Equal Error Rate of 0.0186\%. Furthermore, alternative algorithms for calculating similarity scores are explored to enhance accuracy. Notably, the utilization of a one-class Support Vector Machine reduces the Equal Error Rate to an impressive 0.0163\%. The outcomes of this study indicate that our model surpasses the previous state-of-the-art in free-text keystroke authentication. These findings contribute to advancing the field of keystroke authentication and offer practical implications for secure user verification systems.
\end{abstract}

\section{Introduction}
Keystroke authentication is a type of biometric authentication that relies on the distinct typing patterns of an individual to verify their identity. With online security becoming increasingly crucial and the number of cyber-attacks on the rise, keystroke authentication has emerged as a promising solution to strengthen authentication systems \cite{105}. Compared to traditional authentication methods like passwords and PINs, keystroke authentication offers several benefits. For one, it eliminates the need for individuals to remember complicated passwords, which can be hard to create and recall. Moreover, it provides an extra layer of security that is difficult to duplicate or steal since it is based on an individual's unique typing behavior. Additionally, keystroke authentication is non-intrusive and does not require any physical contact with the user, making it more user-friendly than other forms of biometric authentication. Given its potential to provide a high level of security without requiring additional hardware or software, keystroke authentication has gained significant attention in recent years \cite{50, 124}.

Keystroke identification technology involves analyzing the timing and duration of individual keystrokes, as well as the general layout during a typing session. These patterns depend on several factors, such as the size and shape of the individual's hands, the way they position their fingers on the keyboard, and the speed and rhythm of their typing \cite{103}. By analyzing these patterns, it is possible to construct a profile of an individual's typing behavior, which can be utilized to authenticate them with great accuracy.

There are various applications of keystroke identification, such as controlling access to secure facilities and systems, authenticating online financial transactions, and verifying remote employees' identities \cite{102}. Additionally, the technology can be utilized for continuous authentication, enabling the constant verification of users throughout their session instead of solely during the initial login \cite{52}.

Keystroke authentication can be classified into two main categories: fixed-text and free-text authentication. Fixed-text authentication entails using a predetermined text, such as a password or a passphrase, that the user needs to enter. On the other hand, free-text authentication permits the user to enter any text of their preference. Fixed-text authentication is typically more precise because the user types the same text each time, enabling easier identification of any irregularities or inconsistencies in typing behavior \cite{112}. Conversely, free-text authentication can be more user-friendly as it allows users to choose any text they desire, making it simpler to recall and type quickly.

While keystroke authentication offers numerous benefits, it does have some limitations that need to be addressed. The primary challenge lies in developing algorithms that can accurately analyze an individual's typing behavior, while accounting for variations in patterns caused by factors like stress, fatigue, injury, or keyboard design \cite{122}. Furthermore, privacy concerns arise from the collection and storage of sensitive biometric data \cite{109}. Additionally, keystroke authentication is vulnerable to attacks like replay attacks, where an attacker records an individual's keystrokes and gains unauthorized access \cite{110}. Thus, it is crucial to develop secure and robust keystroke authentication systems that can withstand such attacks.

Despite the challenges, keystroke identification is still a vibrant area of research and development. As the demand for secure and user-friendly authentication methods grows, keystroke identification may play an increasingly pivotal role in the future of biometric authentication. This authentication method can be applied to a broad spectrum of devices, including desktops, laptops, smartphones, and tablets. Moreover, keystroke identification can be integrated with other authentication techniques like passwords and tokens to enhance security \cite{123}.

The literature on keystroke biometrics is vast and contains many promising approaches. However, it is important to acknowledge its limitations. A lot of previous works in keystroke authentication have concentrated on predetermined texts for authentication purposes. This fixed-text approach fails to accurately reflect genuine typing behavior, as users typically type diverse texts of varying lengths \cite{59}. Furthermore, several earlier works have overlooked the importance of scalability in their approaches, relying instead on small datasets derived from limited user populations or contexts. This limitation severely hampers their practical applicability in real-world scenarios, where a vast amount of data is readily accessible \cite{1}.

This paper presents an innovative approach to keystroke authentication that overcomes limitations and substantially improves the accuracy and efficiency of the authentication process. Our approach is specifically designed for free-text keystroke authentication scenarios that more closely reflect real-world typing behavior. To achieve scalability and handle large amounts of data, we utilize a transformer neural network architecture \cite{113}, which has demonstrated excellent performance in processing vast amounts of information. Additionally, our work draws on a comprehensive dataset, meticulously selected to encompass a broad range of users, typing behaviors, and text variations. This extensive dataset enables our model to identify subtle patterns and nuances in individual typing styles, enhancing the accuracy and robustness of our keystroke authentication system. To summarize, our contributions include:
\vspace{-5pt}
\begin{enumerate}
    \item We evaluate the efficacy of two different architectures, specifically the bi-encoder and cross-encoder, in the context of keystroke authentication to determine the architecture that yields the highest effectiveness for this particular task.
    \item In order to enhance the accuracy of our keystroke authentication model, we employ a range of contrastive learning techniques. This entails experimenting with different loss functions and distance metrics, followed by thorough comparisons to determine their effectiveness.
    \item We explore multiple anomaly detection algorithms with the objective of improving the calculation of similarity scores when comparing queries to the enrollment set. Through this investigation, our aim is to enhance the identification of genuine and imposter keystrokes.
    \item Our method attains a significantly lower Equal Error Rate (EER) compared to the previous literature on the widely-known and accessible Aalto keystroke dataset \cite{103}, highlighting the efficiency of our model for the purpose of keystroke authentication.
\end{enumerate}
\vspace{-5pt}
The remainder of this paper is organized as follows: Section \ref{section2} offers an in-depth review of the relevant literature on keystroke biometrics. Section \ref{section3} outlines our proposed methodology, including the transformer architecture and contrastive learning techniques employed. The experimental setup and the dataset used for training and evaluation are outlined in Section \ref{section4}. The results and analysis are provided in Section \ref{section5}. Finally, we conclude our study and delve into future prospects in Section \ref{section6}.

\section{Related Works}
\label{section2}
Keystroke authentication has witnessed significant advancements and contributions throughout its evolution. Monrose and Rubin \cite{22} were pioneers in this field, introducing a groundbreaking algorithm for free-text keystroke authentication. Their approach utilized mean latency and standard deviation of digraphs to compare an unknown input against reference profiles. Building upon their work, Gunetti and Picardi \cite{23} extended the algorithm to n-graphs, further improving its effectiveness.

Subsequent studies have explored different aspects of keystroke biometrics using similar methodologies. For instance, Huang et al. \cite{35} investigated the impact of data size on free-text keystroke performance, while Crawford and Ahmadzadeh \cite{41} examined the influence of user movement during typing on the effectiveness of mobile keystroke dynamics, determining user position before authentication.

Statistical learning algorithms have proven to be highly effective in analyzing keystroke dynamics. Researchers have employed various techniques to model and classify keystroke sequences. Hidden Markov Models (HMM) were used by Jiang et al. \cite{24} to capture the timing information, and this approach was extended to Partially Observable Hidden Markov Models (POHMM) by Monaco and Tappert \cite{44}. In a different study, Ayotte et al. \cite{53} employed a Random Forest (RF) classifier to identify the most significant features in digraph-based algorithms. Additionally, other studies, such as those by Saevanee and Bhatarakosol \cite{25}, and Zahid et al. \cite{27}, demonstrated promising results using k-Nearest Neighbor (KNN) and fuzzy logic, respectively.

Among the various approaches, Support Vector Machine (SVM) has emerged as a popular choice in keystroke biometrics. Ceker and Upadhyaya \cite{38} proposed a combination of the existing digraphs method for feature extraction and an SVM classifier for user authentication. Cilia and Inguanez \cite{49} conducted an extensive study focusing on differentiating typing modes (one or two hands) and user activity (standing or moving) using an SVM-based keystroke verification system. Furthermore, SVM was employed by Gascon et al. \cite{33} in conjunction with mobile sensor data, incorporating the user's motion information while entering text into the smartphone. Regardless of the classifier used, the fusion of keystroke dynamics with simultaneous movement sensor data from mobile devices yielded significant improvements in authentication results \cite{9, 52}.

Benchmark evaluations have played a crucial role in comparing and assessing the performance of different keystroke biometric algorithms. Killourhy and Maxion \cite{26} collected a comprehensive keystroke-dynamics dataset and conducted a thorough comparison of various algorithms. Similarly, Kim et al. \cite{42} performed a benchmark study on algorithms such as Gaussian and Parzen Window Density Estimation, one-class SVM, KNN, and k-means.

In recent years, the field of keystroke biometrics has witnessed remarkable progress in authentication performance, thanks to the emergence of deep learning techniques. Ceker and Upadhyaya \cite{39} explored the applicability of deep learning by utilizing Convolutional Neural Networks (CNN) and Gaussian data augmentation on three diverse datasets. Recurrent Neural Networks (RNN) have also exhibited impressive results \cite{55}, while Multi-Layer Perceptron (MLP) architectures have been extensively explored \cite{58}.

Building upon the applicability of RNNs in capturing temporal patterns, Xiaofeng et al. \cite{50} introduced a fusion of convolutional and recurrent neural networks to extract features for keystroke authentication. RNN variations have also demonstrated their effectiveness in keystroke biometrics. Li et al. \cite{59} introduced a unique method for feature engineering, which involved generating image-like matrices using a hybrid model that combined a CNN with a Gated Recurrent Unit (GRU). While, Acien et al. \cite{1} developed TypeNet, a Siamese Long-Short-Term-Memory (LSTM) model for large-scale keystroke authentication in free-text scenarios. Additionally, Stragapede et al. \cite{3} further advanced the field of keystroke authentication by incorporating a Transformer architecture and leveraging the power of Gaussian range encoding.

Notably, the study by Acien et al. \cite{1} holds particular relevance to this paper, as they employed the same dataset and experimental protocol, achieving state-of-the-art results in free-text desktop keystroke authentication. Thus, their work serves as a significant benchmark for the current study, enabling a direct comparison of the proposed system.

\begin{figure*}[t]
\centering{\includegraphics[width=\textwidth]{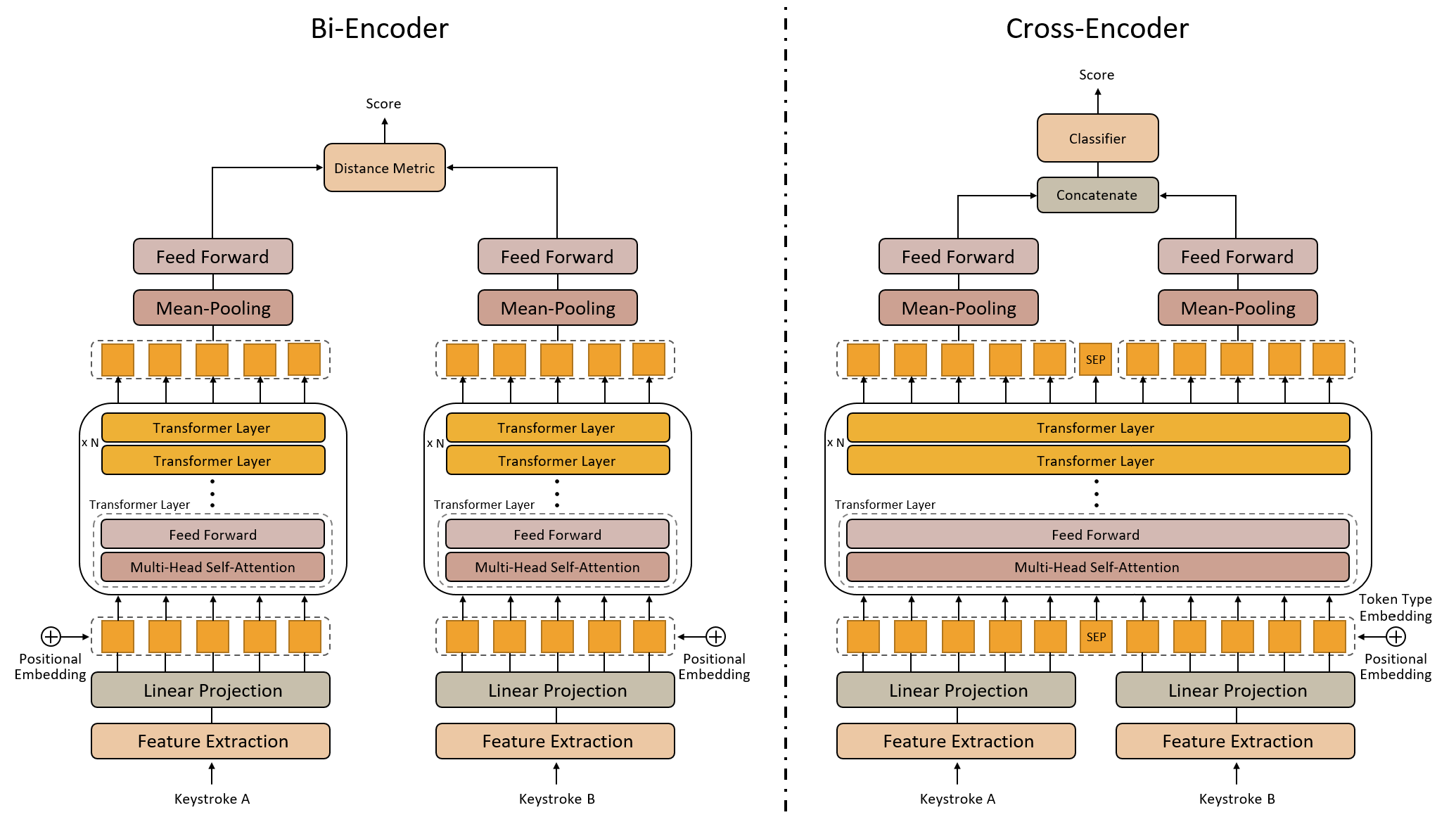 }}
\caption{Comparison between the two proposed architectures: The bi-encoder independently processes each keystroke, generating separate vector representations for each sequence. It utilizes a distance metric to assess the similarity between various sequences. In contrast, the cross-encoder takes both sequences as input and produces an output representation that captures their relationship. This representation is then passed through a classifier to determine the final score.}
\label{fig1}
\end{figure*}

\section{Proposed Method}
\label{section3}
In this study, we focus on evaluating the effectiveness of two different architectures, namely the bi-encoder and cross-encoder, for keystroke authentication. Figure \ref{fig1} presents the pipeline of the proposed architectures. Our goal is to identify the architecture that yields the highest effectiveness for this particular task. To enhance the accuracy of our keystroke authentication model, we employ a comprehensive range of contrastive learning techniques. To begin, we experiment with various loss functions and distance metrics. This allows us to explore different approaches to measure the similarity between keystroke patterns and distinguish genuine users from impostors. By systematically varying the loss functions and distance metrics, we can assess their impact on the performance.

\begin{figure}[h]
\centering{\includegraphics[width=\columnwidth]{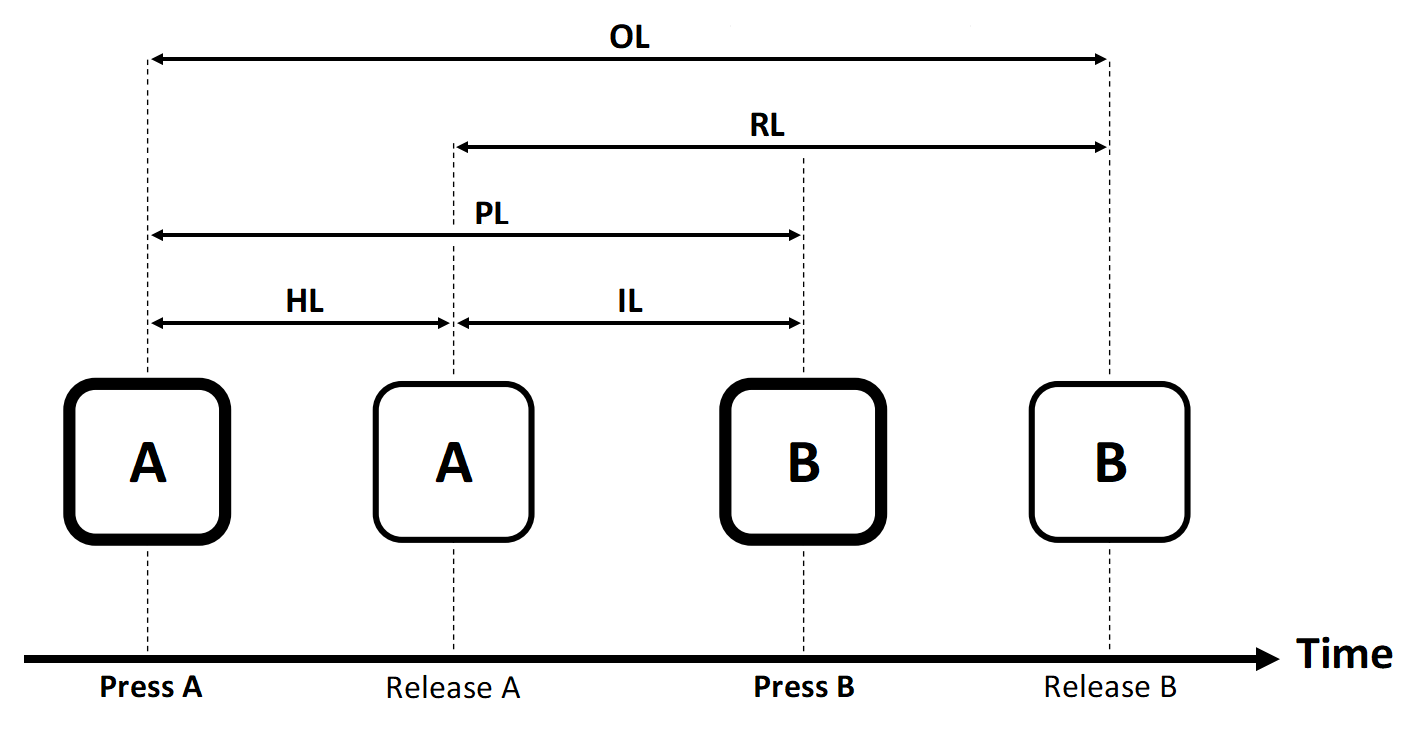 }}
\caption{Example of the distinct temporal features extracted between two successive keystrokes.}
\label{fig2}
\end{figure}

\subsection{Pre-processing \& Feature Extraction}
The raw keystroke data consists mainly of a timestamp indicating when a key is pressed and released, along with the corresponding ASCII code. Keystroke sequences in free-text form can vary in length. In order to maintain a consistent input size, these variable-length sequences are either truncated or padded to a fixed size.

For each sequence, we extract five temporal features: hold latency (HL), press latency (PL), release latency (RL), inner key latency (IL), and outer key latency (OL), as illustrated in Figure \ref{fig2}. These temporal feature values are normalized to have a mean of 0 and a standard deviation of 1 before being used as input for the model. Additionally, we incorporate five spatial features that represent the key positions on the keyboard. Each key is assigned an embedding, which is learned during the training process. The spatial features are derived from the key embeddings of two consecutive keycodes. To achieve this, we apply a one-dimensional convolutional layer with a kernel size of 2 and stride of 1 to the key embeddings matrix. The number of channels in this convolutional layer is set to 5 to match the size of the temporal features. The final input sequence provided to the model has 10 channels, obtained by concatenating the temporal and spatial features.

\subsection{Bi-Encoder}
The bi-encoder model is characterized by its ability to independently process two input sequences and compare them. During the encoding process, there is no direct interaction between the two inputs, and a vector embedding is generated for each sequence. These embeddings capture the essential information from the input data and serve as a basis for calculating a similarity score using a distance metric. This enables the network to effectively determine the similarity between the given input sequences.

To obtain the vector embedding of the keystroke sequence, the extracted features are fed to the bi-encoder. Initially, these features are passed through a linear layer, projecting them to a higher dimension that aligns with the hidden size of the model. In order to preserve positional information during encoding, learnable positional embeddings are incorporated into the sequence. The bi-encoder adopts the widely used transformer architecture, which comprises stacked transformer layers. Each transformer layer consists of a self-attention layer and a feedforward layer each followed by a normalization layer and residual connections. The attention mechanism functions by mapping a set of queries, keys, and values to an output, which is computed as a weighted average of the values. These weights are determined based on the similarity between the key and its corresponding query:
\begin{equation}
\text{Attention}(Q,K,V) = \text{softmax}(\frac{QK^T}{\sqrt{d}})V
\end{equation}
Where $Q$, $K$, and $V$ are the matrices of the queries, keys, and values respectively, and $d$ is the dimension of the input vector. To derive the keys, queries, and values from the input sequence, the self-attention layer utilizes linear projection. Rather than employing a single attention function, each attention module employs multiple heads with distinct parameters. This enables the model to gather information from various representation subspaces:
\begin{equation}
\begin{split}
\text{MultiHead}(Q,K,V) &= \text{Concat}(head_1, \ldots, head_h)W^O \\
\text{where }head_i &= \text{Attention}(QW_i^Q, KW_i^K, VW_i^V)
\end{split}
\end{equation}

Additionally, padded tokens are masked at this stage to prevent them from being attended to by the transformer. Employing self-attention allows the transformer model to efficiently captures dependencies among tokens in the input sequence, enhancing its ability to generate precise and contextually aware representations. Once the sequence has passed through the transformer, the final representation is obtained by applying mean-pooling to the unmasked tokens and passing the resultant vector through a feedforward layer, which further transforms the representation into a suitable space for contrastive learning.

\subsubsection{Distance Metrics}
In contrastive learning, distance metrics are pivotal for assessing the similarity or dissimilarity between sample pairs. These metrics quantify the separation between embeddings in the learned representation space. We utilize three distinct distance metrics:

\textbf{Euclidean Distance:} This fundamental metric calculates the straight-line distance between two points in the embedding space. It is defined as the square root of the sum of squared coordinate differences. Euclidean distance is sensitive to both the magnitude and direction of vector differences.

\textbf{Manhattan Distance:} Also known as L1 distance, this metric measures the distance between two points in a grid-like system. Unlike Euclidean distance, which represents the straight-line distance, Manhattan distance follows a path along the grid's edges.

\textbf{Cosine Similarity:} This metric gauges the cosine of the angle between two vectors, providing a measure of their similarity rather than their distance. It is a common choice in contrastive learning. Cosine similarity is calculated as the dot product of the two vectors divided by the product of their magnitudes. Its range spans from -1 (indicating opposite directions) to 1 (representing parallel directions). Similarly, cosine distance can be defined by subtracting the cosine similarity from 1 and dividing the result by 2, ranging from 0 (for identical vectors) to 1 (for vectors pointing in opposite directions).

\subsubsection{Training \& Loss Function}
To train the bi-encoder model, the embeddings are optimized to minimize the dissimilarity between positive pairs (i.e., pairs of similar sequences) and to maximize the dissimilarity between negative pairs (i.e., pairs of dissimilar sequences). To achieve this, we utilize a contrastive loss function. Specifically, we employ three distinct loss functions in the training process: triplet loss, batch-all triplet loss, and Weighted Decoupled Contrastive Learning (WDCL) loss. Each of these loss functions is defined as follows:

\textbf{Triplet Loss:} The triplet loss function is a common choice for training Siamese networks. It involves the selection of a triplet of samples, including an anchor sample, a positive sample (similar to the anchor), and a negative sample (dissimilar to the anchor). The objective is to maximize the distance between the anchor and negative sequences while minimizing the distance between the anchor and positive sequences. The triplet loss is defined as:
\begin{equation}
\text{Triplet Loss} = \max\{|z_a - z_p| - |z_a - z_n| + \text{margin}, 0\}
\end{equation}
Here, $z$ represents the embedding of the input, $|.|$ denotes the distance function, and the margin is a hyperparameter that controls the separation between positive and negative pairs. The model learns to project similar sequences closer together while pushing dissimilar sequences apart through iteratively sampling and optimizing these triplets.

\textbf{Batch-all Triplet Loss:} In contrast to the triplet loss, which selects a single triplet at a time, the batch-all triplet loss considers all possible triplets within a training batch \cite{118}. It exhaustively evaluates the loss for each triplet, encouraging the model to find the hardest triplets in the batch. This approach ensures that the model optimizes across the entire batch rather than relying on a single triplet at each iteration. The loss is defined as:
\begin{equation}
\text{Batch-all Triplet Loss} = \sum_i\max\{|z_a - z_p^{(i)}| - |z_a - z_n^{(i)}| + \text{margin}, 0\}
\end{equation}
The sum is taken over all possible triplets in the batch, and the max function ensures that only the triplets violating the margin condition contribute to the loss.

\textbf{Weighted Decoupled Contrastive Learning (WDCL):} Decoupled Contrastive Learning (DCL) loss is employed to enhance the discriminative capabilities of the learned representations by decoupling the positive and negative samples during the training process \cite{119}. Traditional contrastive learning encourages positive pairs to be closer together while pushing negative pairs further apart. In large-scale datasets, the number of negative samples can significantly outnumber positive samples, causing an imbalance that can affect the learning process. DCL addresses this issue by decoupling the positive and negative samples in the loss computation, providing more flexibility and control over the learning process. We generalize the loss function to WDCL by introducing a weighting function:
\begin{equation}
\text{WDCL} = - w(z_a, z_p) e^{⟨z_a, z_p⟩} + \log(\sum_i e^{⟨z_a, z_n^{(i)}⟩})
\end{equation}
Here, $w$ is the weight function, $⟨.⟩$ denotes the similarity function, and the sum is taken over all negative samples for the anchor. The weight function can be determined using a function that assigns higher weights to pairs with smaller similarities. We can choose $w$ to be a negative von Mises-Fisher weighting function:
\begin{equation}
w(x, y) = \frac{e^{⟨x, y⟩} / k}{\mathbb{E}[e^{⟨x, y⟩} / k]}
\end{equation}
Where $k$ is a hyperparameter controlling the strength of the weighting. The intuition behind the weight function is that in practice, the data may exhibit varying degrees of similarity, and treating all pairs equally may not be optimal. The weight function addresses this issue by assigning higher weights when a positive pair of samples are far from each other.

\subsection{Cross-Encoder}
In cross-encoder architecture, the input consists of a pair of keystrokes: a source keystroke and a target keystroke. The cross-encoder model takes both sequences as input and encodes them together into a joint representation. It considers the interaction between them and produces a single output representation that captures their relationship.

In order to calculate the similarity score, the cross-encoder utilizes the extracted features from both the source and target keystrokes. Similar to the previous approach, these features are processed through a linear layer, which maps them to a higher dimension equivalent to the hidden size of the transformer. Positional embeddings are introduced to the sequences as well. At this point, the two sequences are padded to enable simultaneous processing by the cross-encoder. Additionally, token type embeddings are incorporated to differentiate between the two sequences within the transformer. The resulting sequence is then passed through the transformer to obtain the joint representation. The cross-encoder structure is similar to the bi-encoder, except that the cross-encoder utilizes self-attention to attend to both sequences. Following the transformer, the source and target sequences are separated and the representation for each sequence is obtained by applying mean-pooling and passing the resulting vector through a feedforward layer. Ultimately, the two representations are concatenated and fed into a linear layer with softmax activation, which converts the joint representation into a two-class probability distribution. The components of this vector represent the probability scores that indicate the level of similarity and dissimilarity between the two sequences.

\subsubsection{Training \& Loss Function}
The cross-encoder model is trained using a supervised learning approach, where the model is provided with a pair of inputs along with a label indicating their similarity or dissimilarity. The model is optimized to predict the correct label for a given keystroke pair. For this purpose, we employ the cross-entropy loss function.

Cross-entropy loss is a commonly used loss function in classification tasks. It measures the dissimilarity between the predicted probability distribution and the true probability distribution of the target variables. In the context of binary classification, where there are two possible classes (0 and 1), the cross-entropy loss can be defined mathematically as follows:
\begin{equation}
\text{Cross\_Entropy}(y, \hat{y}) = -[y \cdot log(\hat{y}) + (1 - y) \cdot log(1 - \hat{y})]
\end{equation}
Here, $y$ represents the true label and $\hat{y}$ denotes the predicted probability. In this equation, the first term $y \cdot \log(\hat{y})$ accounts for the loss when the true label $y$ is 1, while the second term $(1 - y) \cdot \log(1 - \hat{y})$ captures the loss when the true label $y$ is 0.

\section{Experiment Setup}
\label{section4}
\subsection{Dataset}
All the experiments in this paper are conducted using the Aalto desktop keystroke dataset \cite{103}. This dataset is a comprehensive collection of keystroke biometric data obtained from physical keyboards, comprising over 5GB of information gathered from 168,000 participants. To collect the data, participants were instructed to memorize English sentences and type them as quickly and accurately as possible. The sentences were randomly selected from a pool of 1,525 sentences derived from the Enron mobile email corpus and the English Gigaword newswire corpus. The selected sentences range from a minimum of 3 words to a maximum of 70 characters. It is important to note that participants may have made errors while typing, resulting in keystrokes exceeding the 70-character limit as characters could be deleted. Each participant in the Aalto desktop dataset completed 15 sessions, with each session involving typing a single sentence. The captured raw data from each session includes a time series with three dimensions: keycodes, press times, and release times of the keystroke sequence. The timestamps are recorded in UTC format with millisecond precision, and the keycodes range from 0 to 255, corresponding to the ASCII code.

\subsection{Implementation Details}
In order to ensure a fair comparison with previous studies, we followed the protocol introduced by Acien et al. \cite{1} for our training process. We utilized 30,000 subjects from the Aalto dataset exclusively for training purposes. The remaining portion of this dataset was only utilized for evaluating the model, ensuring that there was no overlap between the subjects used for training and evaluation.

We utilize a consistent batch size of 512 across all loss functions. The triplet loss integrates 512 anchors, along with positive and negative pairs, within each batch. Likewise, the batch-all triplet loss employs 64 unique subjects, with 8 samples selected from each subject in every batch. For the WDCL loss batch, we include 512 positive pairs chosen from different subjects, treating each as a negative sample for all other subjects. In the case of the softmax loss, the batch consists of 256 positive pairs and 256 negative pairs. The pairing selection is randomized for all loss functions.

When applying the triplet or batch-all triplet loss, we set the margin to 1 when using the Euclidean and Manhattan distances, and 0.25 when using the cosine distance. Each keycode was represented by a vector embedding of size 16. The transformer architecture consists of 6 layers with a hidden size of 256. We used 8 attention heads and set the intermediate size in the feedforward layers to 512. The transformer incorporates the gelu activation function along with a dropout rate of 0.1. Following the mean-pooling stage, the final feedforward layer reduces the representation size to 64.

All models were implemented in PyTorch, and we employed the Adam optimizer for training. The models underwent a total of 75,000 training steps, starting with an initial learning rate of 0.001 and diminishing by a factor of 0.1 after every 25000 steps, following a step decay schedule.

\subsection{Evaluation Metric}
The Equal Error Rate (EER) serves as a widely accepted metric for evaluating the efficacy of keystroke authentication systems. It delineates the point at which the False Acceptance Rate (FAR) and the False Rejection Rate (FRR) intersect. The FAR pertains to the rate at which impostor (non-matching) samples samples are incorrectly accepted as genuine, while the FRR signifies the rate at which genuine (matching) samples are incorrectly rejected. The computation of the EER entails a comparison of the model's output scores against a predetermined threshold. Should a user's score surpass this threshold, they are accepted; otherwise, they are rejected. Modulating the threshold permits an analysis of the trade-off between FAR and FRR. A lower EER indicates a more precise and dependable authentication system. We investigate two distinct scenarios for EER calculation:

\textbf{Adaptive EER:} This approach involves the selection of an individualized threshold for each subject, resulting in a subject-specific EER value. The final EER value is derived by averaging these subject-specific EER values. The adaptive EER approach bestows the advantage of superior EER performance, as the system tailors a specific threshold for each subject. Nevertheless, a potential drawback is the necessity of a substantial number of samples from each subject to accurately calibrate the threshold. If an insufficient number of samples is available, it becomes challenging to calculate the threshold accurately.

\textbf{Global EER:} In this approach, a single threshold is chosen to be applied universally across all subjects. When the system is trained offline, deploying it with a fixed, pre-determined threshold proves to be more convenient and obviates the need for collecting an extensive number of samples from individual users.

In the present study, we consider both the adaptive EER and global EER scenarios within our experimental framework. By exploring and comparing both approaches, we aim to provide a comprehensive understanding of the system's performance and evaluate the effectiveness of each method.

\begin{table*}[t]
\captionsetup{width=0.77\textwidth}
\centering
\caption{Comparing the effectiveness of different architectures, loss functions, and distance metrics in terms of EER for a keystroke length of 50 characters.}
\label{table1}
\begin{tabular}{c|c|c|cccc}
\hline
\multirow{2}{*}{Architecture} & \multirow{2}{*}{Loss Function} & \multirow{2}{*}{Distance Metric} & \multicolumn{4}{c}{Equal Error Rate (\%)} \\
 &  &  & E = 1 & E = 3 & E = 5 & E = 10 \\ \hline
\multirow{5}{*}{Bi-encoder} & Triplet & Euclidean & 1.6827 & 0.8945 & 0.7596 & 0.6549 \\
 & Triplet & Manhattan & 1.4638 & 0.6782 & 0.5519 & 0.4644 \\
 & Triplet & Cosine & 0.7737 & 0.3587 & 0.2791 & 0.2362 \\
 & WDCL & Cosine & 0.0790 & 0.0314 & 0.0263 & 0.0196 \\
  & Batch-all Triplet & Cosine & \textbf{0.0756} & \textbf{0.0305} & \textbf{0.0249} & \textbf{0.0186}\\ \hline
Cross-encoder & Softmax & - & 0.8521 & 0.3020 & 0.1878 & 0.1372 \\ \hline
\end{tabular}%
\end{table*}

\subsection{Evaluation Protocol}
The authentication process involves comparing a given query with the subject's enrollment set. To evaluate the system's adaptability to varying amounts of enrollment data, we conducted experiments using different configurations of enrollment sessions. Each subject provided a total of 15 keystroke sequences. For our experiments, we randomly selected the enrollment data from the initial 10 samples of each subject, reserving the remaining 5 samples for testing purposes.

Queries can emanate from either the same individual (a genuine match) or a different subject (an impostor match). We treat this as an anomaly detection task to differentiate between impostor and genuine samples. When presented with a query, our objective is to determine whether it corresponds to the subject's enrollment set. To achieve this, we employed various anomaly detection algorithms and assessed their performance.

\textbf{Average Distance:} This method, involving the measurement of the distance between a given query and the enrollment set, proves to be a highly effective and straightforward technique for detecting anomalies. The final score is computed based on the average of these distances.

\textbf{Angle Based Outlier Detection (ABOD):} ABOD stands as a potent technique for detecting anomalous data by assessing their angles in relation to a reference set. It quantifies the angular deviation of a query from the center of the enrollment set, identifying anomalies as queries with significantly distinct angles compared to the enrollment set.

\textbf{Local Outlier Factor (LOF):} LOF is a highly regarded and powerful anomaly detection algorithm. Its primary objective is to evaluate the local density deviation of a query concerning its neighboring data. Intuitively, when a query exhibits a significantly lower density compared to the rest of the enrollment data, it is highly probable to be classified as an anomaly.

\textbf{One-class Support Vector Machine:} One-class SVM is a widely adopted machine learning algorithm for anomaly detection. The basic idea behind a one-class SVM is to create a boundary or hyperplane that encloses the majority of data in a high-dimensional space. The goal is to find a region that maximizes the margin around the enrollment set while minimizing the number of data points outside that region. This region is referred to as the normal region, and any query outside this region is considered an anomaly.

\section{Experiment Results}
\label{section5}
In this section, we present a comprehensive analysis of the results obtained from our experiments, accompanied by an extensive ablation study of our models. Our objective is to delve into the influence of each model component on the overall performance and understand its impact on the task at hand. We provide insights on several aspects, including a comparison between the bi-encoder and cross-encoder architectures, the effects of different loss functions and distance metrics, the impact of enrollment data volume on performance, the effectiveness of anomaly detection algorithms used for similarity score calculation, and a comparison with previous state-of-the-art models.

\subsection{Architecture Comparison}
One of the most critical factors that significantly influences the performance of a neural network is the training process. In this section, we thoroughly examine the impact of different loss functions, distance measures, and model architectures on the overall performance. Among the various loss functions used for training authentication models, the triplet loss stands out as one of the most commonly employed approaches. Therefore, we consider the triplet loss as our baseline and compare the performance of other methods against it. In Table \ref{table1}, we present a comprehensive overview of the adaptive EER exhibited by different methods using varying numbers of enrollment samples (E) for a keystroke length of 50. Notably, the triplet loss attains a EER of 1.682\%, 1.463\%, and 0.773\% when using Euclidean distance, Manhattan distance, and cosine distance, respectively, with only 1 enrollment sample. This outcome emphasizes the significance of the distance function on the model’s performance. Consequently, we opt for the cosine distance to train the remaining models, as it yields superior performance.

Next, we compare the performance of other methods with the triplet loss as a benchmark. For the bi-encoder architecture the EER is 0.0756\% and 0.0790\% for batch-all triplet loss and WDCL loss, respectively, when using 1 enrollment sample. While, the cross-encoder architecture exhibits a higher EER of 0.852\%, contrary to our initial expectations. Typically, cross-encoders tend to outperform bi-encoders due to their ability to leverage both reference and target sequences during the encoding process. However, our findings indicate that the cross-encoder architecture performs acceptably against the triplet loss but falls significantly behind against the batch-all triplet loss and WDCL loss. This leads us to conclude that the superior performance achieved by the batch-all triplet loss and WDCL loss is not solely attributed to the bi-encoder architecture, but rather to the specific loss functions utilized. These loss functions enable the model to derive a more effective representation of the input by comparing multiple samples during training. Without a doubt, the clear winner is the batch-all triplet loss, achieving the lowest EER of 0.0756\% using only 1 enrollment sample. Additionally, due to the utilization of the bi-encoder architecture in this approach, the computational cost is lower compared to the cross-encoder architecture. As a result, the subsequent experiments conducted in this paper exclusively employ the batch-all triplet loss with the bi-encoder architecture as their foundation.

\subsection{Impact of Enrollment Data Quantity}
\begin{figure}[h]
\centering{\includegraphics[width=\columnwidth]{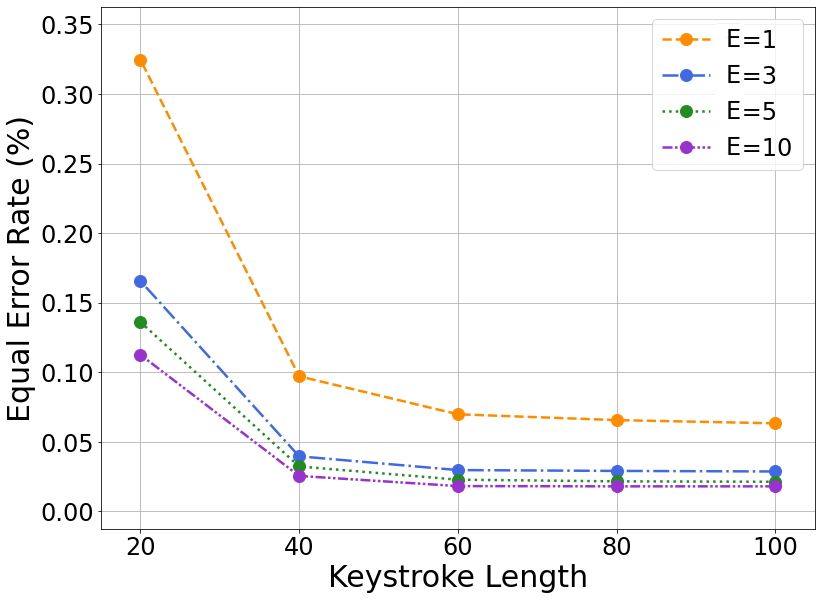 }}
\caption{Influence of enrollment sample count and keystroke length on model's EER}
\label{fig3}
\end{figure}

The performance of keystroke authentication models is significantly affected by the available amount of enrollment data per subject. In this study, we investigate the impact of enrollment data quantity from two perspectives: the number of enrollment samples and the length of each keystroke sequence. Figure \ref{fig3} illustrates adaptive EER of the model across various numbers of enrollment samples and keystroke lengths.

We observed a notable decrease in EER of 0.1148\% when increasing the length of the keystroke sequence from 20 to 100, while utilizing 5 enrollment samples. Within this reduction, a significant improvement of 0.1038\% was observed when transitioning from a sequence length of 20 to 40. However, the improvement gradually diminishes in subsequent stages. A similar trend is evident when considering the number of enrollment samples. Increasing the number of enrollment samples from 1 to 10 resulted in an EER reduction of 0.0570\% for an input sequence length of 50. Notably, a reduction of 0.0451\% was observed when progressing from 1 to 3 samples. In general, increasing the amount of enrollment data enhances the model's performance. The most favorable outcome was achieved by utilizing all 10 enrollment samples with a length of 100, resulting in an EER of 0.0180\%. These findings underscore the importance of having a sufficient amount of enrollment data for accurate keystroke authentication. Longer sequences and a higher number of enrollment samples contribute to improved model performance, ultimately enhancing the effectiveness and reliability of the authentication system.

\subsection{Feature Embeddings Analysis}
A query is accepted in keystroke authentication based on its similarity with enrollment samples. The feature embeddings play a crucial role in this process, as an efficient keystroke authentication model should ensure that samples belonging to the same class (subject) are represented closely together while maintaining a significant distance from samples of other classes.

\begin{figure}[h]
\centering{\includegraphics[width=\columnwidth]{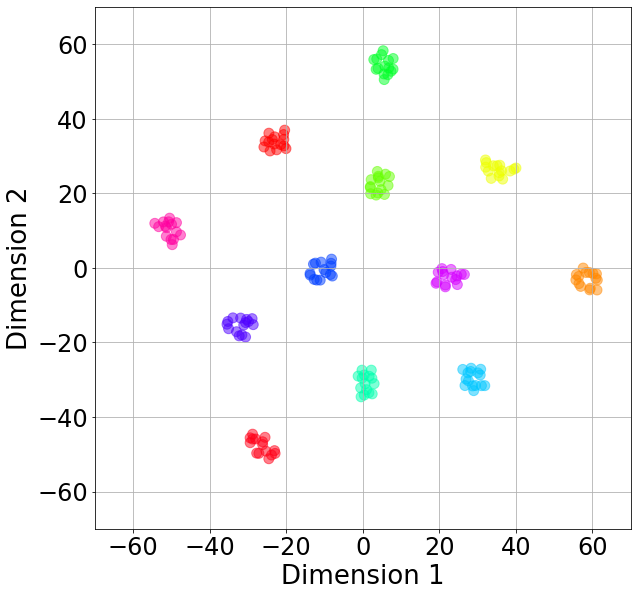 }}
\caption{Feature embeddings derived from keystroke data of various subjects, represented in a 2-dimensional space using t-SNE.}
\label{fig4}
\end{figure}

To evaluate the effectiveness of our approach, we provide feature embeddings of various keystroke samples from 12 distinct subjects. The embedding vectors in our proposed model consist of 64 dimensions. To visualize the distribution of these embeddings, we employ the t-SNE method \cite{120}, which allows us to map the high-dimensional vectors onto a two-dimensional space. The resulting visualization is depicted in Figure \ref{fig4}. Notably, the samples belonging to each subject are distinctly grouped together, exhibiting clear separation from the samples of other classes. This demonstrates the ability of our model to effectively discriminate between different subjects based on their keystroke patterns.

\subsection{Importance of Adaptive Threshold}
\begin{figure}[h]
\centering{\includegraphics[width=\columnwidth]{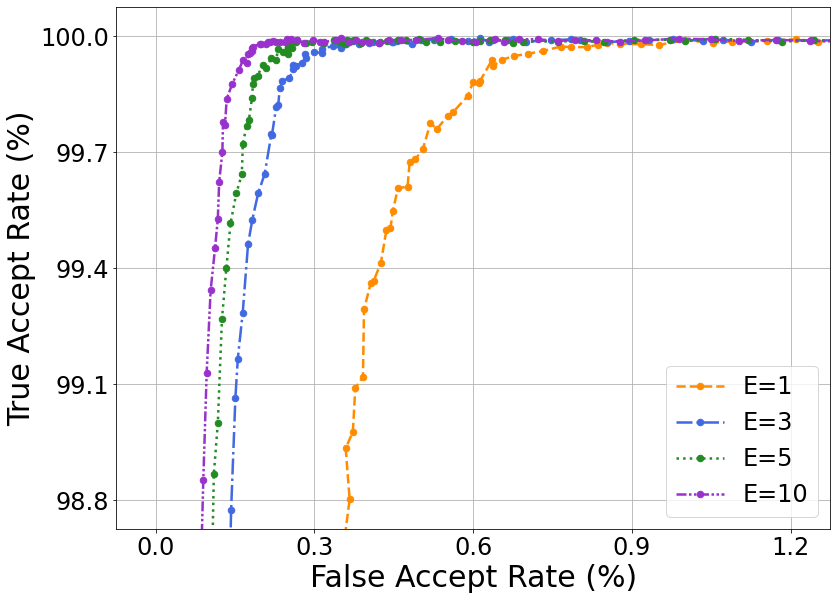 }}
\caption{ROC curve for the global threshold scenario obtained by iteratively changing the threshold value and evaluating the model, with varying enrollment sample sizes and a keystroke length of 50 characters.}
\label{fig5}
\end{figure}

The selection of an appropriate threshold is crucial as it directly impacts both the security and usability of the system. In this section, we will delve into the significance of employing an adaptive threshold and its impact on the performance of the authentication model, comparing it to the use of a global threshold. Figure \ref{fig5} depicts the Receiver Operating Characteristic (ROC) curve obtained when employing a global threshold. This curve demonstrates the relationship between the FAR and the True Accept Rate (TAR) as the threshold of the model is adjusted. The EER corresponds to the point on the curve where it intersects with the diagonal line spanning from the top-left to the bottom-right corners.

\begin{table}[h]
\centering
\caption{Comparing adaptive threshold to global threshold for varying numbers of enrollment data for a keystroke length of 50 characters.}
\label{table2}
\resizebox{\columnwidth}{!}{%
\begin{tabular}{c|cccc}
\hline
\multirow{2}{*}{Scenario} & \multicolumn{4}{c}{Equal Error Rate (\%)} \\
 & E = 1 & E = 3 & E = 5 & E = 10 \\ \hline
Global Threshold & 0.4634 & 0.2407 & 0.1754 & 0.1342 \\
Adaptive Threshold & \textbf{0.0756} & \textbf{0.0305} & \textbf{0.0249} & \textbf{0.0186} \\ \hline
\end{tabular}%
}
\end{table}

A thorough comparison of employing adaptive threshold versus global threshold is presented in Table \ref{table2}. Upon analysis, we discover that by utilizing a global threshold, the model achieves an EER of 0.1342\% using 10 enrollment samples of length 50, which is considerably higher than the adaptive mode's EER of only 0.0186\%. This discrepancy underscores the significance of determining the appropriate threshold for each individual subject. However, it is important to note that in this study, we calculate the adaptive EER by testing different threshold values for each subject. In practical applications, calibrating the threshold for each subject presents numerous challenges. It requires a substantial number of enrollment samples, making it impractical in scenarios where only a few enrollment samples are available per subject.

\subsection{Influence of Scoring Algorithm}
\begin{figure}[h]
\centering{\includegraphics[width=\columnwidth]{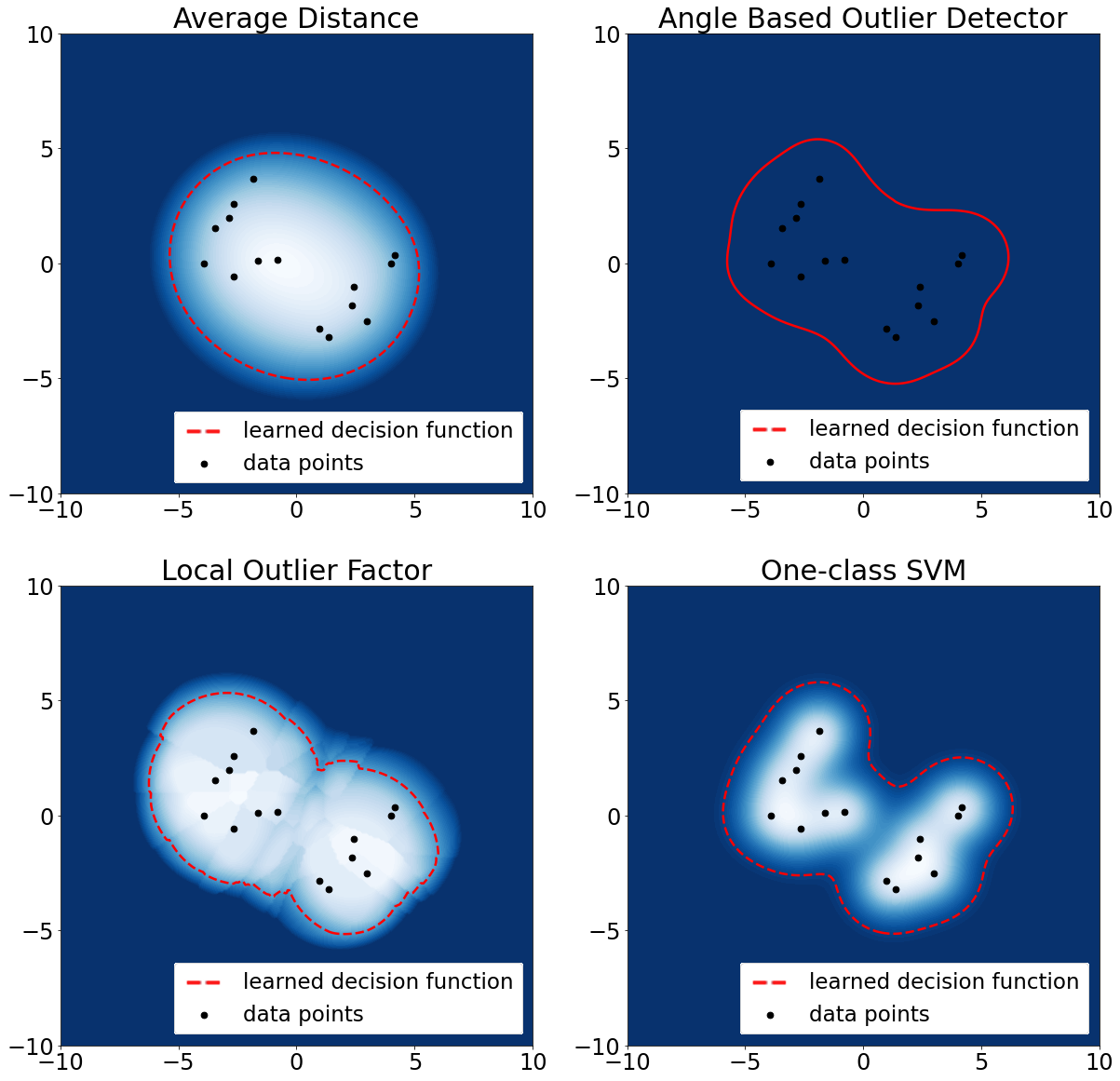 }}
\caption{Illustrating the unique behaviors exhibited by diverse anomaly detection algorithms when applied to a collection of data points.}
\label{fig6}
\end{figure}

The similarity score of a given query is determined by measuring its distance from the enrollment set of a subject. In our approach to calculating this score, we treat the problem as an anomaly detection task. Our aim is to find the most suitable scoring algorithm, and thus, we explore various options and compare their results. Figure \ref{fig6} provides a visual representation of how these algorithms behave when applied to a set of data points.

\begin{table}[h]
\centering
\caption{Comparing the performance of different anomaly detection algorithms with varying numbers of enrollment samples for a keystroke length of 50 characters.}
\label{table3}
\resizebox{\columnwidth}{!}{%
\begin{tabular}{c|cccc}
\hline
\multirow{2}{*}{Method} & \multicolumn{4}{c}{Equal Error Rate (\%)} \\
 & E = 1 & E = 3 & E = 5 & E = 10 \\ \hline
Average Distance & \textbf{0.0756} & 0.0305 & 0.0249 & 0.0186 \\
Angle Based Outlier Detector & - & 0.3826 & 0.0402 & 0.0289 \\
Local Outlier Factor & - & 0.0480 & 0.0274 & 0.0192 \\
One-class SVM & - & \textbf{0.0292} & \textbf{0.0214} & \textbf{0.0163} \\ \hline
\end{tabular}%
}
\end{table}

\begin{table*}[t]
\captionsetup{width=0.715\textwidth}
\centering
\caption{Comparing our proposed model with TypeNet using varying enrollment sample sizes and keystroke lengths.}
\label{table4}
\begin{tabular}{c|c|ccccc}
\hline
\multirow{2}{*}{Model} & \multirow{2}{*}{Keystroke Length} & \multicolumn{5}{c}{Equal Error Rate (\%)} \\
 &  & E = 1 & E = 2 & E = 5 & E = 7 & E = 10 \\ \hline
\multirow{2}{*}{Acien et al. \cite{1}} & L = 50 & 5.4 & 3.6 & 2.2 & 1.8 & 1.6 \\
 & L = 100 & 4.2 & 2.7 & 1.6 & 1.4 & 1.2 \\ \hline
\multirow{2}{*}{Ours} & L = 50 & \textbf{0.0756} & \textbf{0.0389} & \textbf{0.0214} & \textbf{0.0191} & \textbf{0.0163} \\
 & L = 100 & \textbf{0.0633} & \textbf{0.0344} & \textbf{0.0183} & \textbf{0.0165} & \textbf{0.0152} \\ \hline
\end{tabular}%
\end{table*}

For each subject, we train the algorithm using the enrollment set and subsequently predict the similarity score for new query. This score serves as the basis for determining whether to accept or reject different queries. We present the results obtained by each algorithm in Table \ref{table3}. Notably, we observe that both the ABOD and LOF algorithms exhibit inferior performance when compared to simply averaging the distance between the query and enrollment samples. We attribute this decline in performance to the large size of the embedding vector and the limited number of training samples available per subject.

However, amidst these results, we discovered that the one-class SVM algorithm stands out as the most promising one. It not only outperforms the average distance approach but also reduces the EER from 0.0186\% to 0.0163\% when trained on all 10 enrollment samples. Remarkably, even with fewer training samples, this algorithm manages to maintain its strong performance and yields results slightly better than the average distance approach. These findings emphasize the significance of selecting an appropriate scoring algorithm for accurate similarity assessment.

\subsection{Comparison with the State-of-the-art}
In this section, we present a comprehensive comparison between our obtained results and the findings of previous studies. To assess the performance of our proposed model, we evaluate its effectiveness in terms of EER across varying numbers of enrollment samples and keystroke lengths (L), while comparing it to the widely recognized TypeNet model. TypeNet is an LSTM-based architecture that has established itself as a benchmark by achieving state-of-the-art performance on the Aalto keystroke dataset. We selected Typenet for comparison due to its utilization of the same dataset and experimental protocol, providing a fair and direct assessment of our approach. Table \ref{table4} illustrates the performance of our proposed model alongside Typenet. Remarkably, our model outperforms TypeNet by a considerable margin, confirming the efficacy of the presented methodology. These results demonstrate the superior performance of our model, emphasizing its potential for enhancing keystroke authentication.

\section{Conclusion and Future Work}
\label{section6}
In conclusion, this paper presented a comprehensive investigation into keystroke biometrics, with a focus on developing a highly effective system for free-text keystroke authentication. The research explored two transformer-based architectures, the bi-encoder and cross-encoder, and conducted experiments with various loss functions and distance metrics to optimize model training and enhance performance.

The evaluation of the proposed model on the Aalto desktop dataset shows promising results. The combination of a bi-encoder architecture with batch-all triplet loss and cosine distance yields exceptional outcomes. Remarkably, it achieves an EER of 0.0186\% with 10 enrollment samples and maintains a commendable EER of 0.0756\% even with just one enrollment sample. These findings highlight the model's effectiveness in accurately verifying users based on their unique keystroke patterns, making it a reliable tool for free-text keystroke authentication. Furthermore, we explored various algorithms to calculate similarity scores for queries from the enrollment set, including the implementation of a one-class SVM. This approach resulted in an outstanding EER of 0.0163\% with 10 enrollment samples. These achievements present a significant advancement in free-text keystroke authentication, surpassing previous state-of-the-art approaches.

This work provides a highly effective model for free-text keystroke authentication, and the proposed methodologies can serve as a solid foundation for the development of advanced keystroke authentication systems with enhanced accuracy and security. Furthermore, the study's contributions go beyond its model performance by providing valuable insights for researchers and developers in the field. The exploration of various architectures, loss functions, and distance metrics gives a deeper understanding of their impact on performance, enabling others to build upon this work and devise even more efficient keystroke authentication systems.

However, the study also has some limitations. To address them, we suggest that future works explore additional datasets to assess the model's generalization to different populations and diverse scenarios. Additionally, including an analysis of potential adversarial attacks would ensure the proposed system's robustness and security in real-world deployment. Exploring hybrid approaches that combine keystroke biometrics with other authentication methods, such as password-based or behavioral biometrics, could offer enhanced security and usability.

{\small
\bibliographystyle{chicago}
\bibliography{bibliography.bib}
}

\end{document}